\documentclass[aps,prm,twocolumn,raggedbottom,groupedaddress]{revtex4-1}
\usepackage{graphicx}
\usepackage{amsmath}
\usepackage{mathtools}
\usepackage[hidelinks]{hyperref}
\begin{document}

%\preprint{}

\title{Limiting scattering processes in high-mobility InSb quantum wells grown on GaSb buffer systems}


\author{Ch. A. Lehner}
\email[]{lehnerch@phys.ethz.ch}
\affiliation{Laboratory for Solid State Physics, ETH Z\"urich, 8093 Z\"urich, Switzerland}
\author{T. Tschirky}
\affiliation{Laboratory for Solid State Physics, ETH Z\"urich, 8093 Z\"urich, Switzerland}
\author{T. Ihn}
\affiliation{Laboratory for Solid State Physics, ETH Z\"urich, 8093 Z\"urich, Switzerland}
\author{W. Dietsche}
\affiliation{Laboratory for Solid State Physics, ETH Z\"urich, 8093 Z\"urich, Switzerland}
\author{J. Keller}
\altaffiliation{Present address: Institute for Quantum Electronics, ETH Z\"urich, 8093 Z\"urich, Switzerland}
\author{S. F\"alt}
\affiliation{Laboratory for Solid State Physics, ETH Z\"urich, 8093 Z\"urich, Switzerland}
\author{W. Wegscheider}
\affiliation{Laboratory for Solid State Physics, ETH Z\"urich, 8093 Z\"urich, Switzerland}
%\thanks{}


\date{\today}

\begin{abstract}
We present molecular beam epitaxial grown single- and double-side $\delta$-doped InAlSb/InSb quantum wells with varying distances down to 50 nm to the surface on GaSb metamorphic buffers. We analyze the surface morphology as well as the impact of the crystalline quality on the electron transport. Comparing growth on GaSb and GaAs substrates indicates that the structural integrity of our InSb quantum wells is solely determined by the growth conditions at the GaSb/InAlSb transition and the InAlSb barrier growth. The two-dimensional electron gas samples show high mobilities of up to 349\,000 cm$^2$/Vs at cryogenic temperatures and 58\,000 cm$^2$/Vs at room temperature. With the calculated Dingle ratio and a transport lifetime model, ionized impurities predominantly remote from the quantum well are identified as the dominant source of scattering events. The analysis of the well pronounced Shubnikov$-$de Haas oscillations reveals a high spin-orbit coupling with an effective $g$-factor of $-38.4$ in our samples. Along with the smooth surfaces and long mean free paths demonstrated, our InSb quantum wells are increasingly competitive for nanoscale implementations of Majorana mode devices.
\vspace{1cm}
\end{abstract}

%\pacs{}
%\keywords{}

\maketitle


%***************************************** Introduction *************************************************%

\section{Introduction}

High-quality InSb quantum systems are highly desired for the unique and extreme properties of this material in comparison to all other binary III-V compound semiconductors. Aside from applications in high-speed electronics \cite{Ashley1995}, spintronic devices \cite{Zutic2004}, and magnetic sensing \cite{Kazakova2010}, InSb gained a lot of attraction owing to its strong spin-orbit interaction (SOI) \cite{Khodaparast2004} and large Land\'e $g$-factor ($\lvert g \rvert \simeq 51$) \cite{Madelung2000}, in particular for fundamental research in the field of Majorana physics.

%It has been suggested that a semiconductor material exhibiting strong Zeeman splitting in the proximity of a \textit{s}-wave superconductor, may reveal zero-energy Majorana fermion modes \cite{Sau2010, Alicea2010}.
It was suggested that a semiconductor material experiencing \textit{s}-wave superconductivity induced by the proximity effect and exhibiting strong Zeeman splitting hosts zero-energy Majorana fermion modes \cite{Sau2010, Alicea2010}.
Inspired by Oreg \textit{et al.} \cite{Oreg2010} and Lutchyn \textit{et al.} \cite{Lutchyn2010}, the observation of a magnetic field induced zero-bias conductance peak, one of the signatures for a zero-energy Majorana state, was demonstrated in InSb nanowires \cite{Mourik2012, Deng2012}. Quantum computation could be performed in large-scale networks allowing the controlled manipulation of Majorana quasi-particles. However, it is most likely that a top-down patterning approach is the key to pave the way for such a platform for quantum information processing. This favors the implementation of heterostructures with high-quality two-dimensional electron gases (2DEGs) grown by molecular beam epitaxy (MBE).
Recent developments on InAs heterostructures showed that the \textit{in situ} deposition of aluminum as a superconductor in the MBE system can yield electrical transparency between the semiconductor and superconductor \cite{Krogstrup2015, Shabani2016, Kjaergaard2017}, defying diffusive scattering and information loss at the interface, and at the same time indicating stable and strong uniform coupling between the two materials.
Prerequisites to potentially show Majorana zero-energy states are a high crystalline quality in the active region and the absence of residual impurities. In this regard, an accompanying high electron mobility serves as a benchmark indicator. This, in combination with smooth surfaces, creates a basis for transparent interfaces between the semiconductor and superconductor at each point in a possible large-scale network.

In this paper, we show that InSb quantum wells (QWs) grown close to the surface on either [100]-oriented GaAs or GaSb substrates are increasingly competitive for nanoscale fabrications towards Majorana zero-mode favorable devices in terms of electron transport, crystalline structure, and surface morphology. The standard approach in overcoming the large lattice mismatch between InSb and its alloy materials in the III-V semiconductor family, is to adapt an AlSb/InAlSb relaxed metamorphic buffer system on top of a GaAs substrate. However, putting the spotlight on the first transition in these buffers by implementing a GaAs/GaSb transition at the first interface instead, enables us to report on the impact on the morphology and the electronic properties of the 2DEG in a later stage of the growth. The use of GaSb substrates, thereby avoiding one intermediate transition, improves the quality of the samples even further and allows us to draw expediting conclusions regarding the buffer systems used up to date.

In addition, we narrow down the dominant scattering mechanism in our high mobility heterostructures by investigating the quantum scattering time as well as employing a transport lifetime model based on the relaxation time approximation.

%******************************** Adapting optimal growth conditions ***********************************%

\section{Adapting optimal growth conditions for AlSb and GaSb buffer systems}

It is still a great challenge to grow good crystalline quality and high mobility InSb QWs. This is mainly due to the fact, that there are no insulating, high surface quality, lattice-matched InSb substrates available on the market, forcing the community to establish GaAs or GaSb substrates with lattice mismatches of approximately 14.6\% or 6.3\%, respectively, to the QW material instead.
A further issue presents itself with the lack of a lattice-matched barrier material. Typically, an In$_{1-\textrm{x}}$Al$_{\textrm{x}}$Sb barrier with a low Al concentration is chosen to minimize strain in the InSb QW, yet providing sufficient confinement potential.

Although a direct transition from GaAs to InSb, or InAlSb, seems promising and has been investigated by several groups \cite{Zhang1990, Soderstrom1992, Wu2005}, the abundance of threading and misfit dislocations, stacking faults and micro-twins forming at this very GaAs/InSb interface severely affects the integrity of the crystal structure, the surface morphology \cite{Chung2000}, and the carrier mobility \cite{Mishima2005}.
This suggests to implement a metamorphic buffer system allowing a moderate, step-wise adaption of the lattice constant between the substrate and the active region. The main advantages being that a part of the strain can be released much earlier in the system and that an additional distinct transition, the one between AlSb/InAlSb or GaSb/InAlSb, respectively, is capable of releasing strain in an in-planar direction at the interface. This forces already existing micro-twins and threading dislocations out of the crystal and away from the active region or prevents their formation at either interface in the first place \cite{Matthews1976}.

AlSb on GaAs has therefore standardly been utilized as first intermediate buffer layer in most of the investigations on the material system InSb/InAlSb so far. A deeper comparison to GaSb as first intermediate buffer material has yet to be established. In the following, we will adapt virtually the same optimized heterostructure in terms of thicknesses and configuration of the active region to allow a direct and reliable comparison between the two buffer systems. Optimizing the first transition between GaAs and the materials from the 6.1 \AA~family is essential, since it is the first step in shaping and determining the morphology of the subsequent heterostructure. If it is possible to reduce the density of threading dislocations and micro-twins at this stage of the growth, the roughness and scattering potentials can be reduced considerably throughout the QW region and therefore benefit device fabrication on a global as well as on the nanoscale.

All samples reported in this work were grown in a modified Veeco Gen II MBE system with valved cracker cells for the group V component of the compound semiconductors. The protective oxide layers on both, the GaAs and GaSb substrates, were removed by thermal desorption at temperatures of 585$^\circ$C using arsenic and 540$^\circ$C using antimony counter pressure, respectively. To ensure reliable temperature measurement and control during the growth, a pyrometer system with black-body radiation fitting was employed. As a standard, the growth rates were calibrated using reflection high-energy electron diffraction (RHEED) oscillations.
Unless specified, all samples were grown on [100]-oriented, semi-insulating GaAs substrates.

%************************************ First intermediate buffer ****************************************%

\subsection{First intermediate buffer \label{first_buffer}}

\begin{figure}[h]
\includegraphics[width=.45\textwidth]{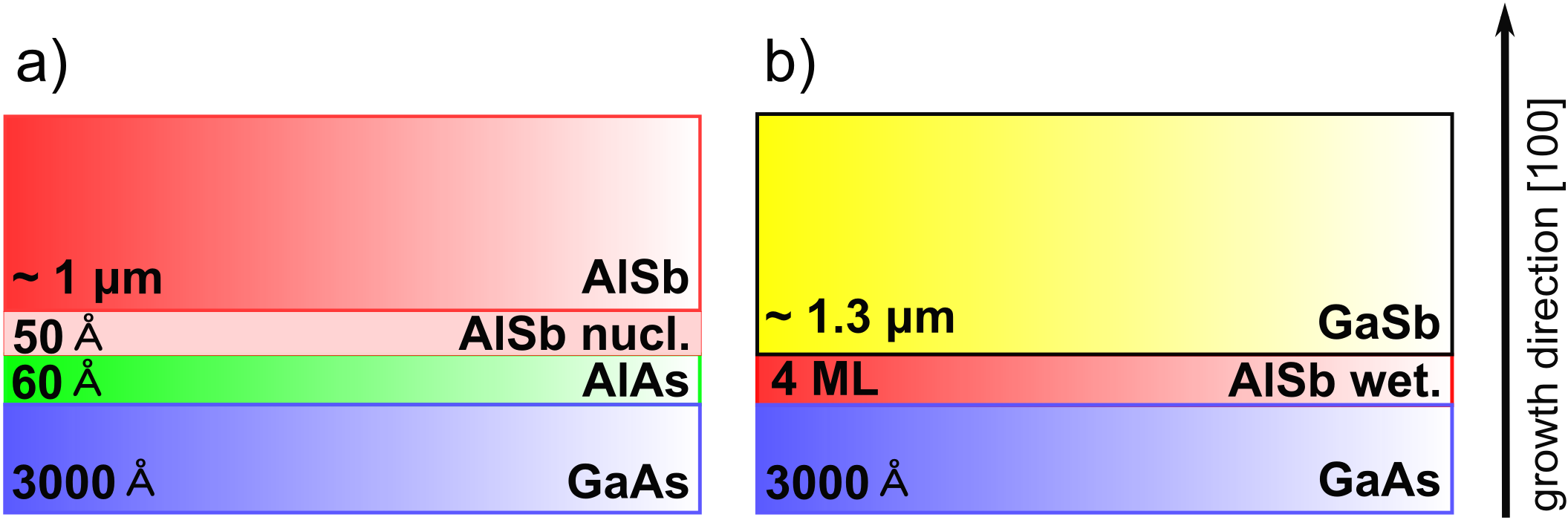}
\caption{Schematic heterostructure of the two first intermediate buffer systems investigated. a) Showing a standard AlSb buffer; b) showing the GaSb interfacial misfit buffer. \label{AlSb-GaSb-transition}}
\end{figure}

Growing the AlSb intermediate buffer, we follow the philosophy of changing only one element for each heterointerface at a time. Therefore, on top of a 3000-\AA-thick GaAs smoothing layer grown at 600$^\circ$C, a nearly lattice-matched thin layer of 60 \AA\, AlAs is grown at the same temperature as schematically depicted in Fig. \ref{AlSb-GaSb-transition}a). Following a 50-\AA-thin AlSb nucleation layer, the roughly 1-$\mu$m-thick AlSb core of the buffer is grown at 540$^\circ$C. This transition sequence is known to exhibit good two-dimensional growth after a short period of time due to the quick nucleation of the AlSb \cite{Subbanna1988, Thomas1997}. However, the temperature and the V/III pressure ratio (Sb/Al) at which the nucleation layer is grown crucially determine the morphology of the subsequent heterostructure.

\begin{figure}[h]
	\centering
	\includegraphics[width=.40\textwidth]{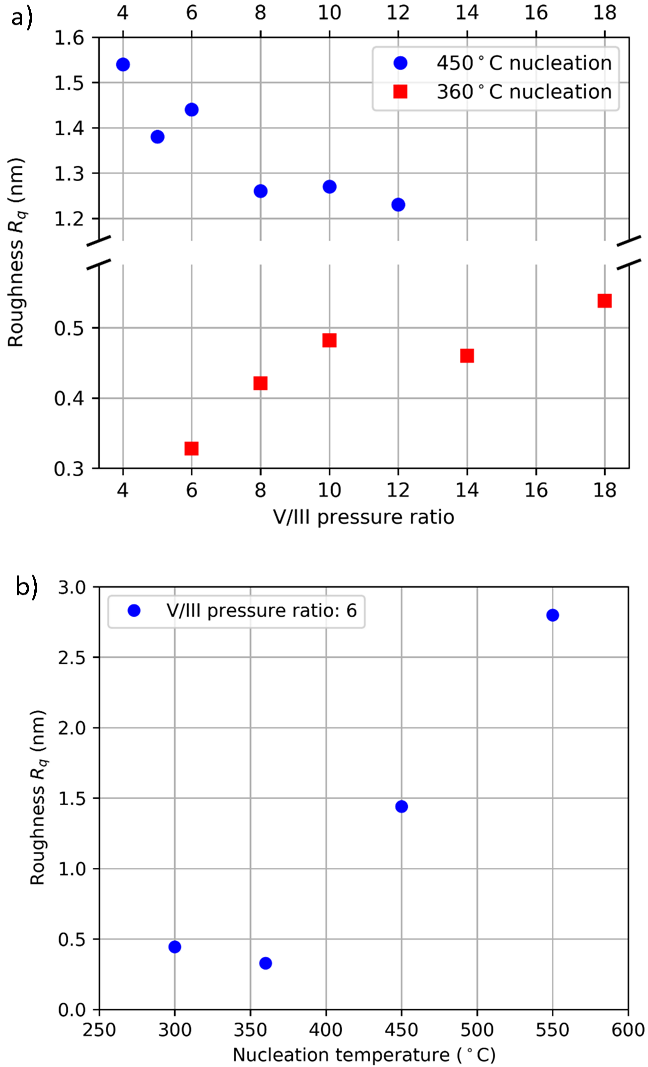}
	\caption{The broken scale plot a) shows the rms roughness $R_q$ in dependence of the V/III pressure ratio for two different AlSb nucleation temperatures. It indicates that a ratio of 6 for a nucleation temperature of 360$^\circ$C yields the lowest roughness at the beginning of the full heterostructure growth. Plot b) shows that for a pressure ratio of 6 the lowest $R_q$ can be achieved with a nucleation temperature of 360$^\circ$C. All values have been measured after 600 nm of AlSb growth and averaged over multiple scans of windows of $5\times 5$ $\mu$m$^2$. \label{fluxes}}
\end{figure}

According to the atomic force microscopy (AFM) data shown in Fig. \ref{fluxes}a), we find a ratio of 6 for a temperature of 360$^\circ$C to be the optimal growth condition at which the major lattice constant transition yields the lowest rms surface roughness $R_q$. The measurements took place roughly 6000 \AA\, into the growth of the AlSb core, which is equipped with a 5 nm GaSb cap to prevent oxidation. This low nucleation temperature and the group V saturation limit allows the newly integrated molecules to slowly adapt the orientation of the homogeneous GaAs/AlAs crystal and at the same time minimizes the formation of possible crystal misfit sites. With these optimized conditions, a multiple scan averaged $R_q$ value as low as 0.328 nm for windows of $5\times 5$ $\mu$m$^2$ can be measured. For a higher nucleation temperature of 450$^\circ$C, the corresponding roughness $R_q$ increases by a factor of 3.75 to 1.23 nm. This illustrates the importance of the nucleation process for the further growth.
In addition, Fig. \ref{fluxes}b) indicates that for an even lower nucleation temperature of 300$^\circ$C at a ratio of 6 the rms roughness starts to increase again, such that a further reduction of the temperature from the optimal 360$^\circ$C turns out to be disadvantageous.

Whereas for the AlSb buffer, in which $60^\circ$ \{111\}-oriented dislocations perforating through the whole heterostructure are predominantly held responsible for the diminishing of the electron mobility \cite{Wen2014, Mishima2003, Mishima2005}, a special growth technique involving a direct transition from GaAs to GaSb is capable of exhibiting purely $90^\circ$ misfit dislocations and allows for an efficient strain relief at this interface \cite{Huang2006, Huang2009}. Applied to a fully grown InSb heterostructure as a standard buffer, this interfacial misfit (IMF) growth technique should ultimately result in a notable reduction of the crystal defect density and, more importantly, in smoother surfaces, causing the electron mobility to rise and the heterostructure to be more suitable for device fabrication.

For the growth of the IMF buffer, we implemented the procedure of Huang \textit{et al.} \cite{Huang2009}. Figure \ref{AlSb-GaSb-transition}b) shows the according schematics. After the desorption of the protective oxide and a 3000 \AA\, GaAs smoothing layer is grown at 600$^\circ$C, the growth is interrupted and the sample cooled down to 540$^\circ$C. Once this temperature is reached, the As valve is closed to ensure a Ga-rich layer forming at the surface due to group V element desorption. As soon as a well pronounced ($4\times2$) RHEED pattern is established, the sample temperature is reduced even further to 480$^\circ$C and the Sb flux is introduced such that the Ga-rich GaAs surface reconstruction changes to ($2\times8$). This ensures that the atomic Sb participates in the formation of a periodic IMF array rather than forming tetragonal distortions in the subsequent layers. Now, 4 monolayers (ML) of AlSb are grown to serve as a wetting layer. For the following roughly 1.3-$\mu$m-thick GaSb layer, the sample temperature is raised to 540$^\circ$C and the characteristic ($1\times3$) surface reconstruction is observed.

%************************** Second intermediate buffer and active region********************************%

\subsection{Second intermediate buffer and active region}

\begin{figure}[h]%
	\includegraphics[width=.49\textwidth]{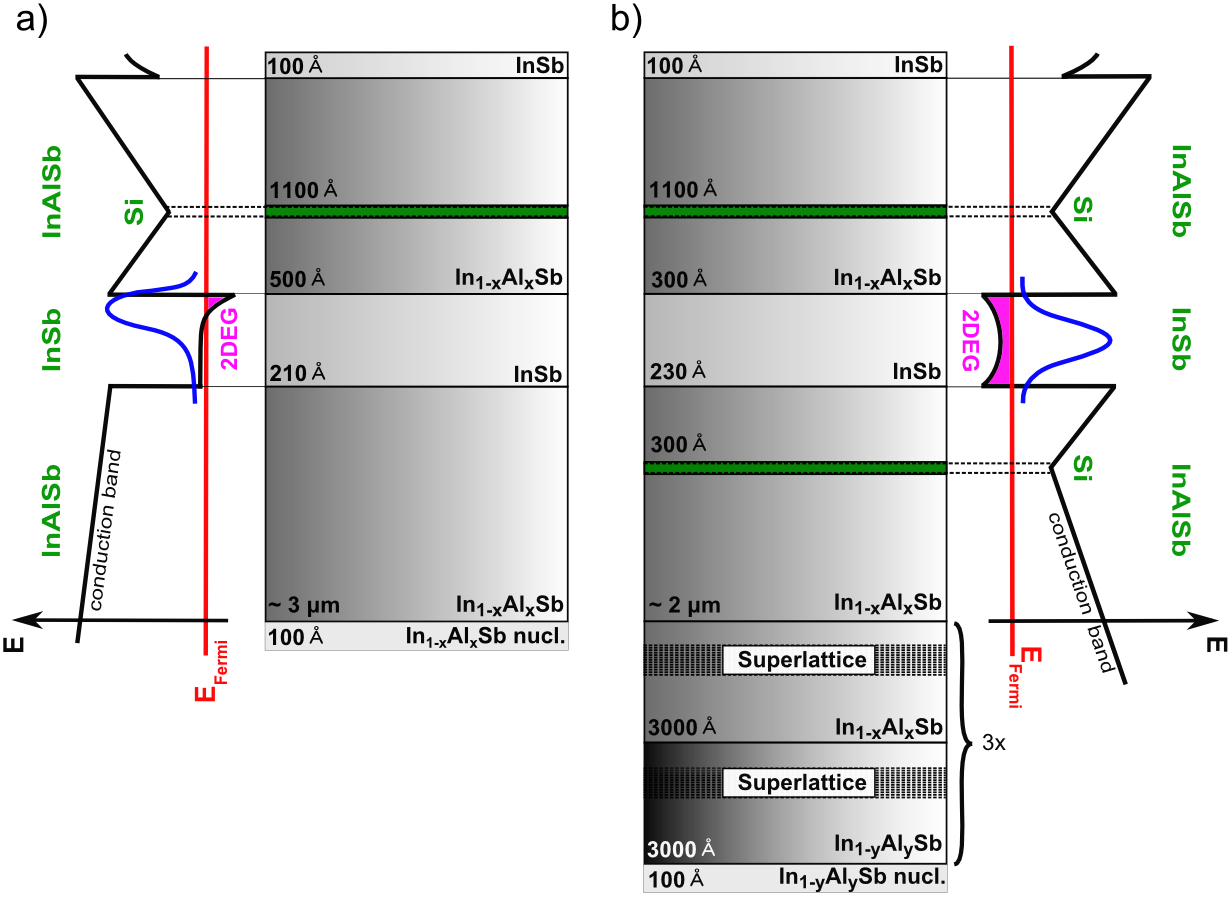}
	\caption{The figure shows the second intermediate buffer and the active region of our samples in a single-side (a) and double-side $\delta$-doped version (b). For both sketches, the conduction band, the 2DEG, and the corresponding first order wave function are provided next to the heterostructure. \label{InAlSb-buffer-transitions}}
\end{figure}

Figure \ref{InAlSb-buffer-transitions} depicts the schematics and layer thicknesses of the two second intermediate buffer systems and active regions considered in this work. In addition, a sketch of the conduction band and the 2DEG including the wave function is displayed.

The InSb QW in the single-side $\delta$-doped (SSD) structure shown in Fig. \ref{InAlSb-buffer-transitions}a) is sandwiched between In$_{1-x}$Al$_x$Sb barriers with an Al content of $x=0.10$. The Si $\delta$ doping resides between the InSb capped surface and the 1700-\AA-deep embedded well. Throughout the whole growth, a substrate temperature of 350$^\circ$C was maintained. With a V/III pressure ratio (Sb/In) of around 2 we observe a c($4\times4$) surface reconstruction, well in agreement with literature \cite{Williams1988,Oliveira1990,Liu1994}. A thin In$_{1-x}$Al$_x$Sb nucleation layer grown at 300$^\circ$C initiates the buffer growth. Since literature \cite{Chung2000,Mishima2004} suggests that InSb grown in a pseudo-($1\times3$) surface reconstruction results in better morphology, we investigated the impact of various growth conditions on the surface roughness roughly 1 $\mu$m into the growth of the In$_{1-x}$Al$_x$Sb buffer for samples with the AlSb buffer outlaid in Fig. \ref{AlSb-GaSb-transition} a). This yields the lowest $R_q$ for the conditions chosen in our system.

The double-side $\delta$-doped (DSD) structure illustrated in Fig. \ref{InAlSb-buffer-transitions}b) has an additional doping layer underneath the QW, such that the conduction band profile between the In$_{1-x}$Al$_x$Sb barriers ($x=0.10$) is symmetric. However, the underlying buffer consists of a three-fold interlayer structure, containing steps of In$_{1-y}$Al$_y$Sb/In$_{1-x}$Al$_x$Sb layers with $y=0.30$. Each InAlSb step has an embedded InSb/InAlSb short-period (25 \AA/25 \AA) strained-layer superlattice. These interlayers \cite{Edirisooriya2007} and superlattices are supposed to contribute to the filtering of threading dislocations. In these DSD heterostructures, we initiate the growth of the second intermediate buffer with an In$_{1-y}$Al$_y$Sb nucleation layer deposited at 300$^\circ$C. In comparison to the SSD structure, the DSD arrangement accounts for the still very large lattice constant difference between the 6.1-\AA\, family and the InSb QW by starting the crystal relaxation with an InAlSb layer of higher Al concentration of 30\%. In principle, this represents an additional step towards the 10\% InAlSb to gradually adapt the lattice constant of the active region.

%****************************** Structural and electrical analysis *************************************%

\section{Structural and transport analysis}
\subsection{Active AlSb and GaSb first intermediate buffer samples}

To establish a well-founded comparison of the effects on the morphology and the electronic properties of an InSb QW heterostructure when adapting the GaSb (sample B) instead of the AlSb (sample A) buffer, the exact same SSD second intermediate buffer in Fig. \ref{InAlSb-buffer-transitions}a) was grown directly on top of the two structures illustrated in Fig. \ref{AlSb-GaSb-transition}.

\begin{figure}[h]
	\centering
	\includegraphics[width=.35\textwidth]{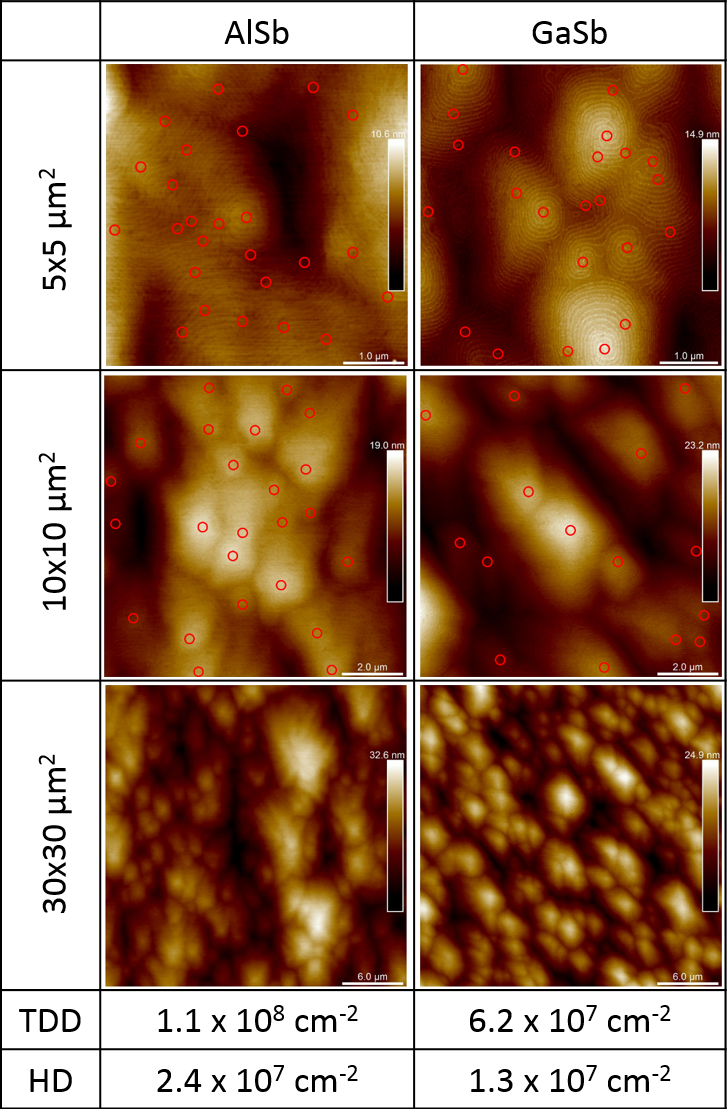}
	\caption{AFM data of the AlSb and GaSb buffers with SSD second intermediate buffer grown on top, revealing the threading dislocation density and the hillock density of the two systems. Red circles in the $5\times5$ $\mu$m$^2$ areas denote threading dislocation sites, whereas in the $10\times10$ $\mu$m$^2$ areas they denote hillocks.}
	\label{Summary_GaSb/AlSb_AFM}
\end{figure}

The AFM analysis of the surface of these samples reveals that the threading dislocation density (TDD) as well as the hillock density (HD) are notably reduced for sample B. While the threading dislocations (TD) and their outcrops \cite{Burton1951}, i.e., merging terraces of single-step atomic planes, undergo a reduction of 43.6\%, the HD is reduced by 45.8\% as demonstrated in Fig. \ref{Summary_GaSb/AlSb_AFM}. With values of $1.1\times10^8$ cm$^{-2}$ (AlSb buffer) and $6.2\times10^7$ cm$^{-2}$ (GaSb buffer) both of these sample types exhibit total defect densities well below typical structures with related buffer systems \cite{Mishima2004} or samples with a direct transition from GaAs to InSb \cite{Jia2015}. Both samples show for highly lattice constant mismatched heterostructures typical spiral growth. Dash-like defects in characteristic crystal directions which would indicate the existence of micro-twins \cite{Mishima2003} are absent in these samples. All the presented defect density values are normalized averages over counts from multiple scans.

\begin{table}[h]
	\caption{Averaged rms roughness $R_q$ (nm) for various window sizes ($\mu$m$^2$) of InSb QWs with either an AlSb or GaSb buffer system. The percentage difference indicates the improvement of the surface roughness in favor of the GaSb buffer samples.}
	\begin{center}
		\begin{tabular}{c|c|c|c}
			\hline\hline
			& AlSb buffer & GaSb buffer & percentage difference\\ \hline
			$R_{q_{(5\times5)}}$		& 3.050 & 2.379 & 22.0\% \\
			$R_{q_{(1\times1)}}$		& 0.558 & 0.417 & 25.3\% \\
			$R_{q_{(0.8\times0.8)}}$	& 0.539 & 0.318 & 41.0\% \\ \hline\hline
		\end{tabular}
		\label{AlSb-GaSb_transition_roughness}
	\end{center}
\end{table}

The reduction of the HD by roughly a factor of 2 from $2.4\times10^7$ cm$^{-2}$ to $1.3\times10^7$ cm$^{-2}$, results in an overall decrease of the surface roughness, for which the values for different sized windows and the percentages denoting the reduction in $R_q$ from the AlSb to the GaSb buffer are displayed in Table \ref{AlSb-GaSb_transition_roughness}. In addition, clear changes in the morphology are visible on large-scale scans of $30\times30$ $\mu$m$^2$. While for the AlSb buffer the hillocks appear unevenly distributed and distinctively differ in size, they are more homogeneously distributed and levelled out in the GaSb buffer samples. This is reflected in the $R_q$ retrieved from areas of $5\times5$ $\mu$m$^2$ where the reduction amounts to 22\%. With $R_q$ being as low as 2.379 nm, the surface roughness is comparable to similar, untreated samples used for device fabrication \cite{Yi2015}.

A standard characterization method for identical van der Pauw square geometries [see inset of Fig. \ref{Magnetoresistance1}a)] at 1.3 K of these heterostructures results in low magnetic field charge carrier densities and electron mobilities of $4.56\times10^{11}$ cm$^{-2}$ and $3.21\times10^{4}$ cm$^2$/Vs for sample A, as well as $3.05\times10^{11}$ cm$^{-2}$ and $7.54\times10^{4}$ cm$^2$/Vs for sample B, respectively. This denotes a major increase by a factor of more than 2.3 in the electron mobility in favor of the GaSb buffer sample, which is well in agreement with the observed reduction in the TDD, the HD, and the sample roughness with an enhanced sample quality therefrom. It cannot be concluded that the electron density difference between samples A and B is solely due to the different buffers, and thus defect densities, since similar density variations are observed in samples with comparable overall defect densities (see below, samples C and D).

All improvements in the crystal structure and electrical transport can be attributed to the growth dynamics of AlSb and GaSb \cite{Thomas1997} in the first intermediate buffer. The different surface mobility of the Ga and Al adatoms during growth can result in a distinctly different morphology of a sample. While Al adatoms tend to get incorporated into the crystal approximately at sites where they hit the sample surface, Ga may show a higher surface mobility and therefore makes a smoothing of the surface during growth possible \cite{Brar1995,Blank1996}. Hence, AlSb shows a tendency to conserve the morphology of underlying epilayers and will pass along the roughness initially created at the transition interface. In contrast, GaSb allows for less distortion of the QW caused by roughness from the interface, yielding higher electron mobility in the channel. The results confirm the importance of the careful optimization of the first intermediate buffer for InSb QW heterostructures.

%*************************************** High performance QWs *******************************************%

\subsection{High-quality InSb quantum wells grown on GaSb buffers and substrates}

High-mobility InSb QWs are achieved when growing optimized SSD (sample C) and DSD (sample E) InSb QWs combining the IMF transition buffer with the three-fold interlayer buffer illustrated in Fig. \ref{InAlSb-buffer-transitions}b), where in the SSD case we skip the lower doping layer and use a QW width of 210 \AA. Moreover, we draw a comparison of sample C to an exact copy of this second intermediate buffer grown on a [100] GaSb substrate (sample D), for which after a 3000 \AA\, smoothing layer of GaSb deposited at a substrate temperature of 540$^\circ$C, only one grown transition is necessary to reach the lattice constant of In$_{1-y}$Al$_y$Sb. Table \ref{high-mob_comparison} displays the relevant data of the analysis, i.e., the TDD, the HD, $R_q$, as well as the charge carrier density $n_e$ and the electron mobility $\mu$. 

\begin{table}[h]
\caption{Characteristics of the GaSb buffer SSD (sample C), DSD (sample E), and GaSb substrate SSD (sample D) InSb QWs. The charge carrier density $n_e$ and electron mobility $\mu$ measurements were performed on square samples using the van der Pauw technique at 1.3 K.}
\begin{center}
\begin{tabular}{c|c|c|c}
	\hline\hline
	Sample & C & D & E \\ \hline
	TDD ($10^7$ cm$^{-2}$) & 13.8 & 9.2 & 9.2  \\
	HD ($10^7$ cm$^{-2}$) & 3.8 & 3.8 & 1.4  \\
	$R_{q_{(5\times5)}}$ (nm) & 3.461 & 2.844 & 3.434 \\
	$n_e$ ($10^{11}$ cm$^{-2}$) & 2.85 & 4.02 & 4.90 \\
	$\mu$ ($10^5$ cm$^2$/Vs) & 2.17 & 2.41 & 3.49 \\ \hline\hline
\end{tabular}
\label{high-mob_comparison}
\end{center}
\end{table}

\begin{figure*}
	\centering
	\includegraphics[width=0.9\textwidth]{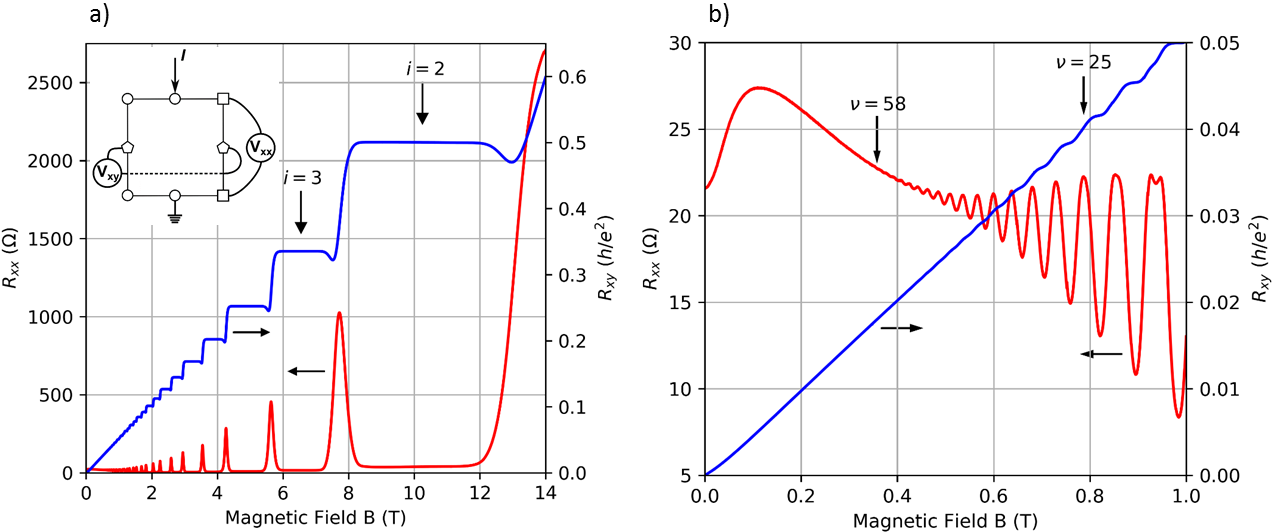}%
	\caption{a) Magnetoresistance data measured at 250 mK of sample F, demonstrating the integer quantum Hall effect up to fields of 14 T. b) Low-field region of the longitudinal ($R_{xx}$) and transverse ($R_{xy}$) resistance, indicating the onsets of the SdH oscillation at an even filling factor of $\nu=58$ and the Zeeman splitting at an odd $\nu=25$. The inset in a) shows the schematics of the van der Pauw measurements performed on all the samples.}
	\label{Magnetoresistance1}
\end{figure*}

The overall TDD and HD are slightly larger when comparing to the IMF sample B presented in the previous chapter. This stands in contrast to the filtering effect from the built-in interlayers and superlattices which should lead to a decrease in the defect density \cite{Edirisooriya2007,Shi2017} and is not observed in our samples. Sample C shows distinctive dash-like slip lines along the [110] and [1$\bar{1}$0] directions which are related to micro-twins \cite{Mishima2003,Mishima2004}, resulting in a higher TDD.
The increased defect densities can be explained by the considerably smaller thickness of the In$_{1-x}$Al$_{x}$Sb lower barrier implemented here opposed to the case in Fig. \ref{InAlSb-buffer-transitions}a). This results in a reduced probability for self-annihilation of threading dislocations in the second intermediate buffer over the course of the shorter growth. Moreover, it is possible that the transition from GaSb to In$_{1-y}$Al$_{y}$Sb presented here and the interlayer compound are capable of offering additional TD nucleation sites, increasing the overall defect density. Nonetheless, the presented samples still show defect densities of lower order than previously reported \cite{Mishima2004,Jia2015}.

The defect densities between samples C, D, and E are roughly equal. Assuming the GaSb substrate to be defect free, this result strongly indicates that the main source of defects in our samples is governed by growth conditions at the transition from the first to the second intermediate buffer and confirms the effectiveness of the IMF transition for InSb QW heterostructures. 

Using a GaSb rather than a GaAs substrate yields a 17.8\% reduced surface roughness $R_q$ of 2.844 nm. In addition, the mobility is increased to 241\,000 cm$^2$/Vs at a density of $4.02\times10^{11}$ cm$^{-2}$, which we mainly attribute to the decrease in $R_q$. The reduction in the surface roughness can prove to be of vital importance for device processing, where flat surfaces are sought for.

The observed substantial jump in the mobility in comparison to samples A and B may additionally be accounted for the application of a higher Al concentration, where Al being highly reactive can act as a getter of charged background impurities. A larger amount of these impurities can then be incorporated deeper in the buffer, such that their scattering potentials are out of range for the active region. This effect can be observed in GaAs/AlGaAs heterostructures grown in our group. It suggests that the interlayers with higher Al concentration of 30\% in our InAlSb metamorphic buffers act as charged background impurity traps.

With the additional $\delta$-doping layer in the lower buffer for sample E, it is possible to compensate for electrons originating from the doping layer residing above the QW which saturate surface states instead of populating the well. In addition, the wave function of electrons occupying the lowest energy state is drawn towards the center of the QW, which decreases the sensitivity of the carriers towards the interfaces of the square well, as illustrated in the band alignment in Fig. \ref{InAlSb-buffer-transitions}b). Thus, the carrier density in sample E is enhanced and it shows a high mobility of 349\,000 cm$^2$/Vs, being only short to a Te SSD single QW \cite{Gilbertson2009} with an ungated mobility of $\mu=$ 395\,000 cm$^2$/Vs at a density of $3.28\times10^{11}$ cm$^{-2}$.

\begin{figure}[h]
	\includegraphics[width=.45\textwidth]{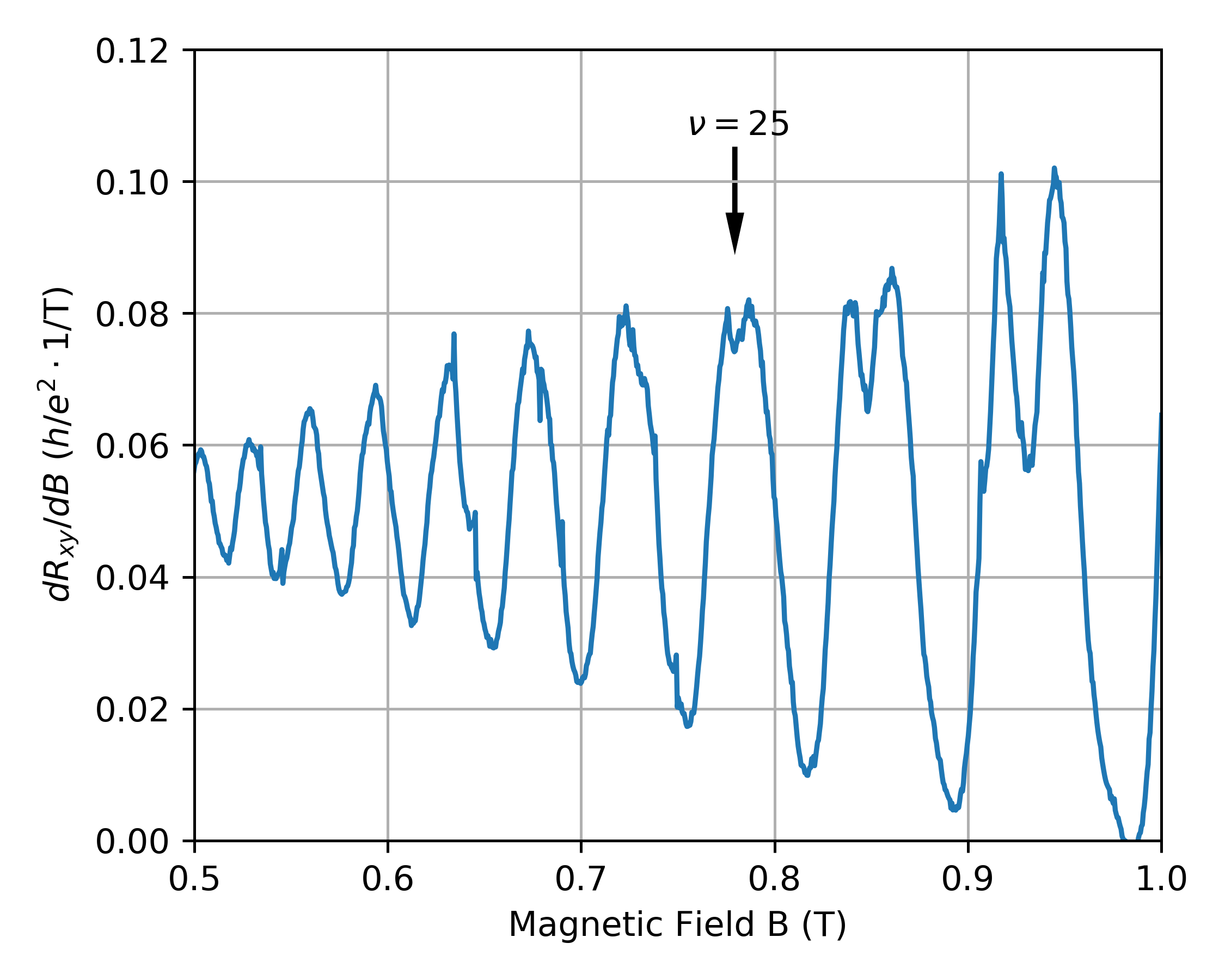}
	\caption{First derivative of the measured Hall resistance $R_{xy}$ in sample F with respect to the magnetic field $B$. It reveals the onset of the Zeeman splitting at a magnetic field corresponding to $\nu=25$. \label{derivative}}
\end{figure}

Figure \ref{Magnetoresistance1} shows the transport data measured at $T=250$ mK for a van der Pauw square geometry [see inset Fig. \ref{Magnetoresistance1}a)] of a sample (F) similar to sample E with a density of $4.53\times10^{11}$ cm$^{-2}$ and mobility of 314\,000 cm$^2$/Vs. Clear quantization plateaus in the Hall resistivity arise at values of $R_{xy}=h/ie^2$ ($i=1,2,...$) for magnetic fields $\mu B \gg 1$. Displayed by arrows in Fig. \ref{Magnetoresistance1}a) are the second and third Landau levels. The respective minima in the longitudinal resistivity $R_{xx}$ indicate single-subband occupation and exhibit single-period oscillations. Figure \ref{Magnetoresistance1}b) shows the low magnetic field region, in which the Shubnikov$-$de Haas (SdH) oscillations are resolved up to an even filling factor of $\nu=n_eh/eB=58$. The onset of Zeeman spin splitting at an odd $\nu=25$ (determined by the first derivative of the Hall resistance as shown in Fig. \ref{derivative} is indicative of these samples exhibiting a large Land\'e $g$-factor and corroborates the high quality of the InSb QWs.

To get a reasonable estimation of the effective $g$-factor $g^*$ of the 2D electrons in sample F, a closer investigation on the role of the Landau level broadening $\Gamma$ is needed. If the thermal contribution of $\Gamma$ is insignificant and since the oscillations in $R_{xx}$ manifest oscillations of the density of states, it can be assumed that these become resolved at a critical field $B_{c_1}$ for which the cyclotron energy exceeds $\Gamma$, hence the Landau level separation gets visible. Therefore, the level broadening is simply given by $\Gamma = \hbar eB_{c_1}/m^*$. Similarly reasoned, a critical field $B_{c_2}$ defines the onset for which the energy $\Delta E_z$ related to the Zeeman spin splitting exceeds $\Gamma$. If the level broadening is field independent between the two critical fields, the energies can be set equal to give $\hbar eB_{c_1}/m^* = \mu_Bg^*B_{c_2}$. With the Bohr magneton $\mu_B=e\hbar/2m_e$ a simple expression for the effective $g$-factor is then given by
\begin{equation}
g^* = \frac{2m_eB_{c_1}}{m^*B_{c_2}}\,.
\label{g-factor}
\end{equation}
The effective electron mass $m^*/m_e=0.0248$ in sample F has been determined by cyclotron resonance measurements, of which the results are illustrated in the Appendix. Applied to Eq. (\ref{g-factor}) and retrieving the critical fields from the transport data ($B_{c_1} = 0.371\pm0.002$ T and $B_{c_2} = 0.779\pm0.005$ T) leads to a very high $\lvert g^* \rvert$ of $38.4\pm0.5$ for narrow well InSb heterostructures, which exceeds values for structures of similar design \cite{Nedniyom2009}. The field independence of $\Gamma$ in our samples will be discussed in Sec. IV.B.

We note that this method \cite{Zverev2004} and the outcome of the calculation crucially depends on the exact determination of the critical fields $B_{c_1}$ and $B_{c_2}$, which in turn is done by carefully examining the first and second derivatives of $R_{xx}(B)$ versus $1/B$ as well as the first derivative of the Hall resistance as shown in Fig. \ref{derivative}. In addition, the value of $g^*$ for narrow QW structures is influenced by strain in the QW and the exchange interaction of the electron wave function with the barrier \cite{Moeller2003}, as well as the degree of non-parabolicity \cite{Krishtopenko2012} of the subbands expressed through the effective electron mass. Further influence may arise form spin polarization \cite{Nedniyom2009}, which manifests itself in an exchange enhanced spin splitting of the individual Landau levels and therefore in an increased $g^*$. However, for 2DEGs exerting large filling factors (as is the case for sample F) the contribution of the polarized states to the total system $\xi$ and with that the broadening of the Landau levels due to exchange interaction becomes negligible and the effective $g$-factor is represented by the polarization-independent, ``bare" $g_0$. The simple picture described above should nonetheless be treated as an approximation, rather than an exact determination of $g^*$.

%************************************* Scattering and Discussion ***************************************%

\section{Scattering Mechanisms}

\subsection{Influence of crystal defects on electron transport}

In devices especially designed for the detection of Majorana fermions with the capability of providing a platform for the controlled braiding manipulation of the quasi-particles, minimal disorder and therefore ballistic transport of the electrons within the charge carrying material is one of the key features \cite{Fadaly2017}. Not only does this require the charge carrying material to exhibit high crystalline quality with low defect densities, but also for it to provide a 2D landscape in which the scattering potentials minimally affect the electron transport. Thus, two characteristic length scales, namely, the Fermi wavelength $\lambda_F = \sqrt{2\pi/n_e}$ and the electron mean free path \cite{Hirayama1991}
\begin{equation}
l_e = \sqrt{\frac{2\pi\hbar^2n_e\mu^2}{e^2}} \,
\label{meanfreepath}
\end{equation}
provide insight into the interplay between defects and electron scattering. Table \ref{mean_free_path} lists $\lambda_F$ and $l_e$ for samples A through E, as well as the average distance between the TDs $d_{\textrm{TD}}$ and the average extent $L_H$ in distinct crystal directions of the pyramid-like hillocks retrieved from the AFM analysis.

\begin{table}[h]
\caption{The electron mean free path $l_e$ and Fermi wavelength $\lambda_F$ determined by the electron mobility and density for samples A through E. $d_{\textrm{TD}}$ denotes the average distance between the threading dislocations and $L_H$ the average extension of the hillocks retrieved from the AFM analysis.}
\begin{center}
\begin{tabular}{c|c|c|c|c}
	\hline\hline
	Sample & $l_e$ ($\mu$m) & $\lambda_F$ (nm) & $d_{\textrm{TD}}$ ($\mu$m) & $L_H$ ($\mu$m) \\ \hline
	A & 0.358 & 37.1 & 0.953 & - \\
	B & 0.683 & 45.4 & 1.270 & - \\
	C & 2.025 & 46.0 & 0.758 & 2.585 \\
	D & 2.521 & 39.5 & 1.043 & 2.087 \\
	E & 4.031 & 35.8 & 1.043 & 2.254 \\ \hline\hline
\end{tabular}
\label{mean_free_path}
\end{center}
\end{table}

The electron mean free paths in our samples are considerably shorter (samples A and B) and longer (samples C through E) than the corresponding distances $d_{\textrm{TD}}$. This means that for the latter samples the contribution of the TDs in our large-scale 2D samples does not limit the elastic mean free path. In addition, it is mostly scattering potentials varying on the scale of the Fermi wavelength (short-range scatterers) which affects electron transport in the QW strongly. Because $\lambda_F$ varies from 35.8 to 46.0 nm and the possible morphological scattering centers in the horizontal dimension to the well are more than 2 $\mu$m in size, which is roughly 60 times $\lambda_F$, the hillocks act as a slowly varying, large-scale scattering potential for the electrons. This indicates that the TDs and the hillocks \cite{Thomas1997} in our samples have a negligible effect on the transport properties of the InSb QW.

Moreover, these results are highly interesting in the prospect of Majorana physics specialized device fabrication. If brought closer to the surface, the 2DEGs performance and the potential of offering electrically transparent surfaces, will not be affected by the defects in the crystal but more by the actual layer sequence and therefore the engineering of the heterostructure and the energy bands themselves, such that the electron wave function may interact with a suitable superconducting material grown on top of the sample \cite{Shabani2016}. In fact, for an optimized inverted SSD sample G (the characteristic data are displayed in Table \ref{invertedsample}) we can show that the morphology of the sample does not affect the 2DEG. Sample G was grown on a GaAs substrate with GaSb first intermediate buffer [Fig. \ref{AlSb-GaSb-transition}b)] and a similar second intermediate buffer to the one in Fig. \ref{InAlSb-buffer-transitions}b), employing solely the lower doping layer and neglecting the InSb cap. The 21 nm InSb QW resides 50 nm beneath the surface. With $l_e = 854$ nm being clearly shorter than the average distance $d_{\textrm{TD}}$ between the TDs and the Fermi wavelength $\lambda_F = 51.8$ nm being a factor 46.5 smaller than the average hillock size $L_H$, these possible scattering potentials are again insignificant for the electron transport in the sample, hence, the very high mobility of 107\,000 cm$^2$/Vs for a 2DEG close to the surface. The electron mean free path in sample G is considerably larger than in typical high-mobility InSb nanowires \cite{Plissard2012,Weperen2013,Gazibegovic2017}. We note here that in windows with a side length in the order of $l_e$, sample G shows a very low rms roughness of 0.679 nm. The results motivate that with the optimized growth conditions presented in this work, InSb QWs are highly competitive to the InSb nanowires and eligible for large-scale Majorana networks.

\begin{table}[h]
\caption{Characteristics of an inverted SSD sample with the 2DEG residing 50 nm below the surface.}
\begin{center}
\begin{tabular}{c|c}
	\hline\hline
	Sample & G \\ \hline
	$n_e$ & $2.34\times10^{11}$ cm$^{-2}$ \\
	$\mu$ & $1.07\times10^5$ cm$^2$/Vs \\
	$l_e$ & 0.854 $\mu$m \\
	$\lambda_F$ & 51.8 nm \\
	$d_{\textrm{TD}}$ & 1.043 $\mu$m \\
	$L_H$ & 2.409 $\mu$m \\
	$R_{q_{(5\times5)}}$ & 3.399 nm \\
	$R_{q_{(0.85\times0.85)}}$ & 0.679 nm \\ \hline\hline
\end{tabular}
\label{invertedsample}
\end{center}
\end{table}

\subsection{Quantum scattering lifetimes}

Considering the minor influence of the crystal defects on the electron transport in our heterostructures, other scattering mechanisms must be responsible for limiting the electron mobility. The quantum lifetime $\tau_q$ can expose the long- or short-range nature of the dominant scattering potentials in the structure when combined with the transport lifetime $\tau_{tr}$. It describes the average time a charge carrier remains in a particular momentum eigenstate when scattering is present (momentum relaxation) and, for small magnetic fields, is related to disorder-induced broadening of the Landau levels by\cite{DasSarma1985,Coleridge1989} $\Gamma = \hbar/2\tau_q$. 

\begin{figure}[h]
	\centering
	\includegraphics[width=.45\textwidth]{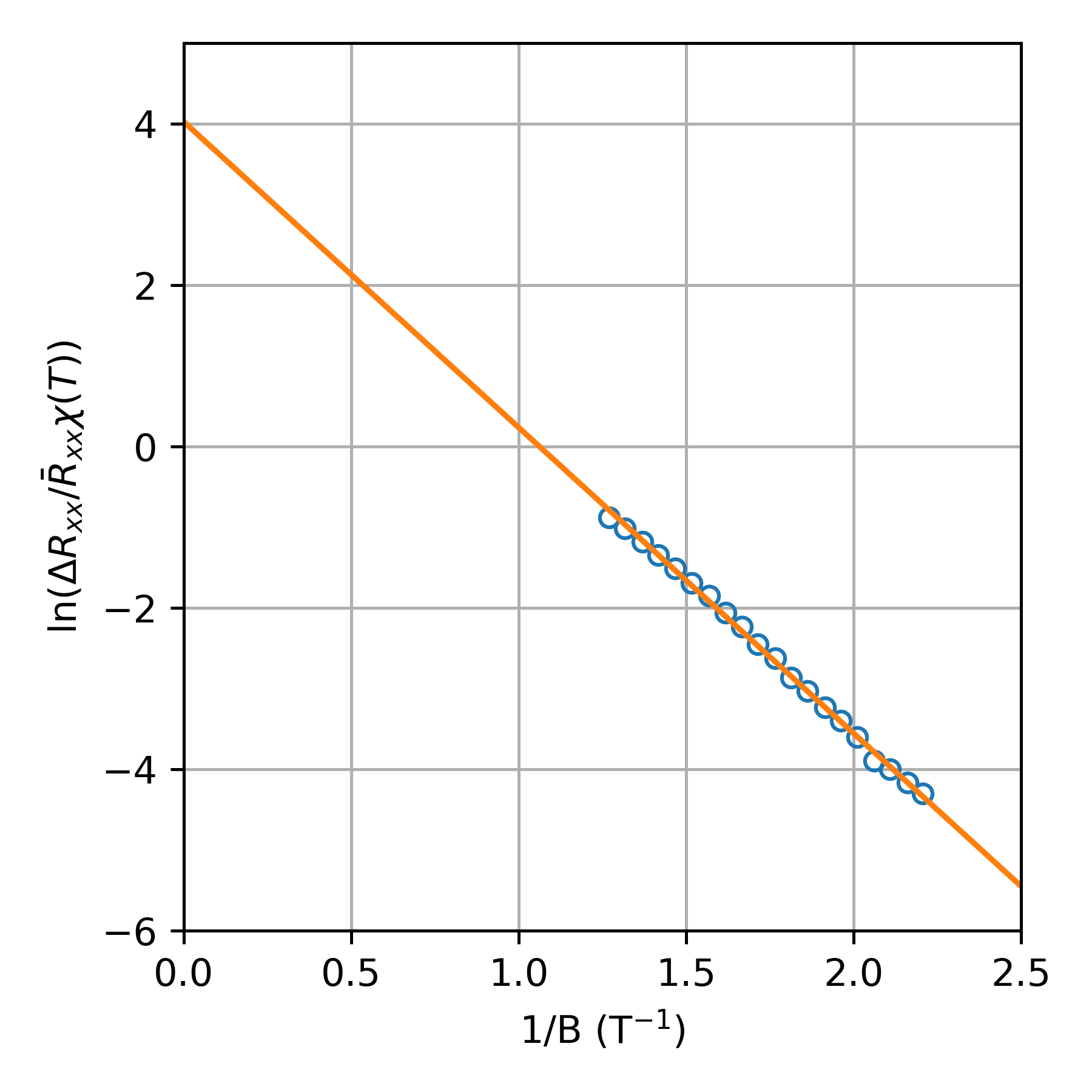}
	\caption{Dingle plot generated from the analysis of the transport data observed in Fig. \ref{Magnetoresistance1}.}
	\label{dingle}
\end{figure}

At any given temperature $T$, $\tau_q$ is extracted from Dingle plots \cite{Dingle1952}. The slope of these plots determines $1/\tau_q$ and can be deduced experimentally from the envelope of the oscillations in the longitudinal resistivity $\Delta R_{xx}/\bar{R}_{xx}$ by \cite{Coleridge1991,Ihn2010}
\begin{equation}
\frac{\Delta R_{xx}}{\bar{R}_{xx}} = 4X(T)\text{exp}\left(-\frac{\pi}{\omega_c\tau_q}\right)\,,
\end{equation}
where $\bar{R}_{xx}$ is the non-oscillatory background resistivity, $\omega_c = eB/m^*$ is the cyclotron frequency, and $X(T)=(2\pi^2 k_BT/\hbar\omega_c)/\text{sinh}(2\pi^2 k_BT/\hbar\omega_c)$ is the thermal damping factor. The result of the analysis of sample F is shown in Fig. \ref{dingle}, where the logarithm of the envelope of the evaluated data from Fig. \ref{Magnetoresistance1} divided by $X(T)$ is plotted against the reciprocal magnetic field.
The linearity of our calculated data seen in the Dingle plot, is indicative that the Landau level broadening in sample F is field-independent \cite{Zverev2004} between the onset of the oscillations and the spin splitting, which supports the calculation of the effective $g$-factor $g^*$ previously expounded. The results also portend a valid fit \cite{Coleridge1991} for $\tau_q$ since the intercept at $1/B=0$ is 4. This additionally confirms that no adversely acting parallel conduction is present, which is supported by the observed transport data. Analyzing the Dingle plot gives the quantum lifetime $\tau_q = 0.12$ ps for sample F.

An indication of the dominant angle of scattering in the 2DEG is given by the Dingle ratio \cite{DasSarma1985,Harrang1985,Coleridge1991} $\tau_{tr}/\tau_q$, where we define $\tau_{tr}$ in the next chapter. While $\tau_{tr}$ is weighted towards large-angle scattering, $\tau_q$ reflects scattering mechanisms influenced by small and large angles. Consequently, if the Dingle ratio is close to 1, isotropic short-range scattering is expected. The ratio can be considerably larger than 1, if the dominant scattering mechanism originates from long-range Coulomb interactions.

With $\tau_{tr} = 4.43$ ps for sample F the Dingle ratio is $\tau_{tr}/\tau_q = 37.9$, which is comparable to high-mobility GaAs and SiGe quantum wells \cite{Coleridge1991,Ismail1995}, as well as to high-mobility InAs channels \cite{Tschirky2017} at similar electron densities. This suggests the dominant scattering mechanism to predominantly originate from ionized impurities remote from the QW, exerting long-range Coulomb interaction leading to small scattering angles.

\subsection{Remaining scattering mechanisms}

To disclose the remaining relevant scattering mechanisms, a transport lifetime model can give further insight in their relative ascendancy. In the transport formalism of the relaxation time approximation, the averaged transport lifetime $\tau_{tr}$ determines the electron mobility by \cite{Ihn2010}
\begin{equation}
\mu_e = \frac{\left|e\right|\tau_{tr}}{m^*} \,.
\label{mobility}
\end{equation}
It incorporates the disorder in the system influencing the electron transport within the QW caused by various scattering mechanisms, of which the most prominent ones are phonon scattering ($\tau_{op}$ and $\tau_{ac}$), in which perturbations in the crystal potential affecting the electrons are described, and charged impurity scattering, which attributes the influence of remote ($\tau_{rii}$) and charged background impurity ($\tau_{cbi}$) potentials. Interface roughness scattering ($\tau_{int}$), which is exclusive to heterostructure systems, accounts for roughness at the interface between the QW and the barrier material, which can lead to a change in energy of the electron wave packet and is strongly dependent on the width of the QW. For heterostructures with ternary barrier material, the mobility can additionally be diminished by alloy scattering ($\tau_{alloy}$). All these mechanisms contribute to $\tau_{tr}$ with their respective scattering rates as reflected in Matthiessen's rule
\begin{equation}
\frac{1}{\tau_{tr}} = \sum_{i}\frac{1}{\tau_i} = \frac{1}{\tau_{op}}+\frac{1}{\tau_{ac}}+\frac{1}{\tau_{rii}}+\frac{1}{\tau_{cbi}}+\frac{1}{\tau_{int}}+\frac{1}{\tau_{alloy}}\,.
\label{1/tautr}
\end{equation}

In the following outline, we assume that intersubband scattering is absent. For a QW of width $w_0$, scattering by polar optical phonon absorption of quantized electrons bound to the well yields a characteristic scattering rate \cite{Price1981,Ridley1982}
\begin{equation}
\frac{1}{\tau_{op}} = \frac{e^2\omega_0N(\omega_0)m^{*2}w_0}{4\pi\epsilon_p\hbar^2}\,,
\end{equation}
where $\epsilon_p^{-1}=\epsilon_\infty^{-1}-\epsilon_s^{-1}$ with $\epsilon_\infty$ and $\epsilon_s$ being the high-frequency and static dielectric constants, $\hbar\omega_0$ is the optical phonon energy, and with the Boltzmann constant $k_B$,
\begin{equation}
N(\omega_0) = \left( e^{\frac{\hbar\omega_0}{k_BT}}-1 \right)^{-1}
\end{equation}
gives the number of phonons.

Acoustic phonon scattering is determined by the deformation potential constant $\Xi$, the crystal density $\rho_d$, and the longitudinal sound velocity $v_s$. The scattering time for acoustic phonons then is \cite{Arora1985}
\begin{equation}
\frac{1}{\tau_{ac}} = \frac{3 m^* \Xi^2 k_B T}{2 \hbar^3 \rho_d v_s^2 w_0}\,.
\end{equation}

Within the Born approximation, the remaining individual momentum relaxation times for a two-dimensional system are given by the Stern-Howard formula \cite{Stern1967}
\begin{align}
\frac{\hbar}{\tau_i} &= \frac{1}{2\pi\varepsilon_F}\int_{0}^{2k_F} \frac{q^2}{\sqrt{4{k_F}^2-q^2}}\frac{\langle\vert U_i(q)^2 \vert\rangle}{{\epsilon(q)}^2} dq\\
&=\frac{m^*}{2\pi\hbar^2}\int_{0}^{2\pi}\frac{\langle\vert U_i(q_\varphi)^2 \vert\rangle}{{\epsilon(q_\varphi)}^2}\left(1-\text{cos}(\varphi)\right) d\varphi\,,
\label{tau_phi}
\end{align}
where $\varepsilon_F$ is the Fermi energy, $k_F$ is the Fermi wave number, $\epsilon(q)$ is the dielectric matrix and $q_\varphi=\vert\vec{k}-\vec{k'}\vert=k\sqrt{2(1-\text{cos}(\varphi))}$. Since the electrons contributing to conduction are close to the Fermi energy, the wave vector $k$ can be set equal to $k_F = \sqrt{2\pi n_s}$ with the sheet carrier density $n_s$. We include the screening effect through the Thomas-Fermi approximation\cite{Ihn2010} such that $\epsilon(q)=1+q_{\textrm{TF}}/q$, where $q_{\textrm{TF}}=2/a_B^*$ denotes the Thomas-Fermi wave vector and $a_B^*=a_B\frac{m_e}{m^*}$ is the effective Bohr radius of the 2DEG carrying material. The averaged squared matrix element of the random potential $\langle\vert U_i(q)^2 \vert\rangle$ is specific to the individual forms of disorder \cite{Gold1988}.

We account for the symmetry of our square QWs by using the symmetric wave function $\psi(z)$ in growth direction as
\begin{equation}
\psi(z) = \left(\frac{2}{w_0}\right)^{1/2}\text{sin}\left(\frac{\pi z}{w_0}\right),\, 0 \leq z \leq w_0
\end{equation}
and zero for all other $z$. This leads to the squared matrix element of the random potential induced by remote ionized impurities located at a spacer distance $d$ to the well of
\begin{equation}
\langle\vert U_{rii}(q) \vert^2\rangle = n_d \left(\frac{e^2}{2\epsilon_s\epsilon_0q}\right)^2 F_{rii}(q,z_i)^2\,.
\end{equation}
The form factor comprises the finite thickness of the problem with
\begin{align}
F_{rii}(q,z_i) &= \int_{-\infty}^{+\infty}\vert\psi(z)\vert^2e^{-q\vert z-z_i\vert} dz\\
&=\frac{4\pi^2}{q w_0}\frac{1-e^{-qw_0}}{4\pi^2+q^2 w_0^2} e^{-qz_i}\,
\end{align}
and satisfies $F_{rii}(q\rightarrow0)=1$. Here, $n_d$ denotes the density of ionized impurities in the doping plane, $z_i$ accounts for the distance between the impurity layer and the QW, and $\epsilon_s$ is the static dielectric constant of the host material.

For homogeneously distributed charged background impurities in the QW with a 3D density $N_B$, the random potential takes the form \cite{Gold1987}
\begin{equation}
\langle\vert U_{cbi}(q) \vert^2\rangle = N_Bw_0\left(\frac{e^2}{2\epsilon_s\epsilon_0q}\right)^2F_{cbi}(q)\,,
\end{equation}
with
\begin{equation}
F_{cbi}(q) = \frac{1}{w_0}\int_{-\infty}^{+\infty} dz_i F(q,z_i)^2\,.
\end{equation}

The surface roughness of the interface between the barrier and the QW can be described by the function $f(x,y)=f(\vec{r}) \eqqcolon \Delta(\vec{r})$, such that the QW width $w(\vec{r})=w_0+\Delta(\vec{r})$ fluctuates, where $w_0$ denotes the intended width of the QW after growth. For the assumption of zero vertical electric field \cite{Jana2011} and of infinitely high barriers to the quantum well, the ground-state energy $E_0 = \pi^2\hbar^2/2m^*w_0^2$ varies as \cite{Sakaki1987}
\begin{equation}
\delta E_0(\vec{r}) = \frac{\partial E_0(0)}{\partial w(\vec{r})}\Delta(\vec{r}) = -\frac{\pi^2\hbar^2}{m^*w_0^3}\Delta(\vec{r}) \eqqcolon F_{int} \Delta(\vec{r})\,.
\end{equation}
This results in the random potential for interface roughness scattering of the form
\begin{align}
\langle\vert U_{int}(\vec{\rho})\vert^2\rangle &= \bigg\langle\frac{1}{A}\bigg\vert\int d^2r \, F_{int} \Delta(\vec{r}) e^{i\vec{\rho}\vec{r}} \bigg\vert^2\bigg\rangle\\
= \frac{1}{A} F_{int} & \int d^2r d^2r' \langle \Delta(\vec{r})\Delta(\vec{r}') \rangle e^{i\vec{\rho}\vec{r}} e^{i\vec{\rho}\vec{r}'} \,,
\end{align}
in which $A$ is a normalizing area for the two-dimensional integral. Since in our samples only the statistical properties of $\Delta(\vec{r})$ are of importance, we chose for the fluctuation correlation $\langle \Delta(\vec{r})\Delta(0) \rangle = \Delta^2 \text{exp}(-r^2/\Lambda^2)$ with the correlation length $\Lambda$ as outlined in Prange \textit{et al.} \cite{Prange1968}. With this correlation and the Fourier transform of the integral, the random potential then results in \cite{Jana2011}
\begin{equation}
\langle\vert U_{int}(q)\vert^2\rangle = \pi F_{int}^2 \Delta^2 \Lambda^2 e^{-\frac{q^2\Lambda^2}{4}}\,.
\label{U_int}
\end{equation}

Alloy disorder (AD) is characterized by the strength parameter $V_{\textrm{AD}}$ and the unit cell $a^3$ of the ternary alloy by \cite{Ando1982,Gold1988,Gold2013}
\begin{equation}
\langle\vert U_{\textrm{AD}}(q)\vert^2\rangle = x(1-x)\frac{a^3}{4}V_{\textrm{AD}}^2F_{\textrm{AD}}\,,
\end{equation}
where $F_{\textrm{AD}} = \int dz \vert\psi(z)\vert^4$ is the AD form factor.

The temperature dependence incorporated in the formalism enables us to determine the dominant scattering mechanism within the heterostructure at a given temperature range and therefore allows us to gain insight into possible adjustments needed in structure engineering to further increase the mobility of the active channel, especially at cryogenic temperatures.

The intrinsic material parameters used for this work are \cite{Dixon1980,Shao2007} $\epsilon_s = 16.52$, $\epsilon_\infty = 15.7$, $\rho_d=5790$ kg/m$^3$, $v_s=3700$ m/s, and $\hbar\omega_0=25$ meV. Furthermore, with the effective mass $m^*=0.0254\,m_e$ retrieved from cyclotron resonance measurements (see Appendix), $a_B^*=35$ nm and $q_{\textrm{TF}}=0.057$ nm$^{-1}$. The sheet density is noted in Table \ref{high-mob_comparison}. The depletion plane is $d=30$ nm away from the $w_0=23$ nm QW and we estimate the density to be $n_d=4.41\times10^{12}$ cm$^{-2}$. We account for 
%the incorporation of charged background impurities earlier in the growth (see previous chapter) and
the low Al content of 10\% in the barrier of our samples by adapting the value $N_B=1.5\times10^{13}$ cm$^{-3}$, which is reasonable compared to literature \cite{Nott2000,Orr2008}. The deformation potential $\Xi$ for InSb has been studied extensively \cite{Tukioka1991} and shows a large variation in absolute value, strongly depending on the experimental method applied. With the lowered \cite{Orr2008} Al concentration in the barrier, the deformation should, in principle, be small, such that we chose $\Xi=7.2$ eV. From the AFM analysis of samples similar to sample E we find the average variation at the interface in $z$ direction $\Delta=4.2$ \AA\, (step height at the interface), as well as the measured roughness correlation length $L=\Lambda=48.2$ nm, which is in good agreement with literature \cite{Hong1986,Noda1991,Bolognesi1992} on heterostructures with morphologically similar sample surfaces. $F_{\textrm{AD}}$ has been calculated from $8\times8$ k$\cdot$p simulations, in which only the part of the wave function entering the barrier has been taken into account.

\begin{figure}[h]
	\centering
	\includegraphics[width=.45\textwidth]{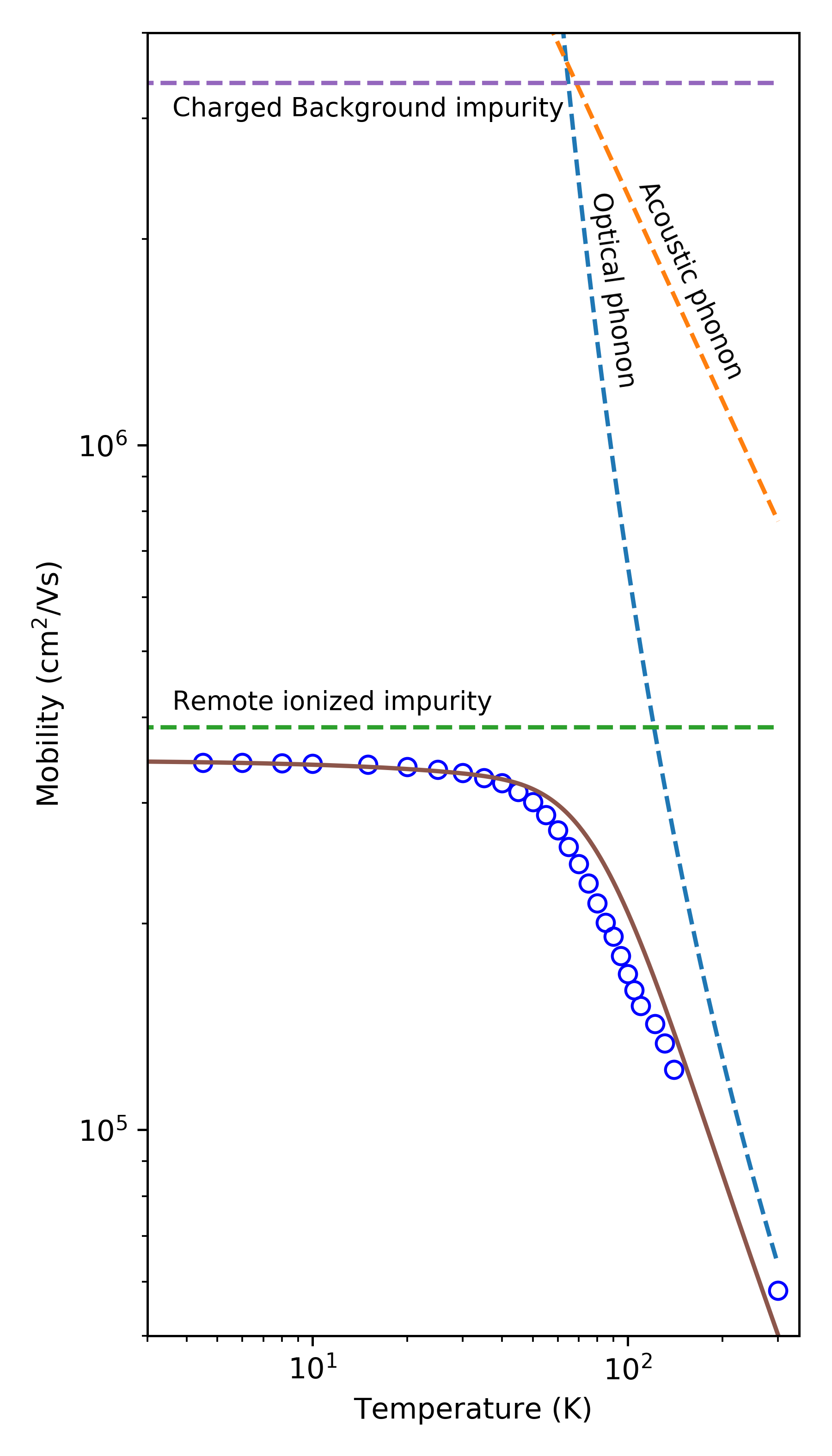}
	\caption{Temperature-dependent mobility (open cirles) for a 23 nm InSb QW heterostructure with a 30 nm spacer (sample E) and the fit (solid line) from the described transport lifetime model. The dashed lines show the calculated mobility limitations from the individual scattering mechanisms.}
	\label{mobplot}
\end{figure}

Figure \ref{mobplot} shows the temperature-dependent mobility measurement of sample E, as well as the mobility limiting curves from Eqs. (\ref{mobility}) and (\ref{1/tautr}). The fit represented by the solid curve agrees remarkably well with the measured data. From the analysis we find that the temperature dependence of the charge carrier mobility enters through scattering on lattice vibrations. At low temperatures the vibrations freeze out and the mobility is limited by remote ionized impurity scattering. This result is supported by the calculations of the Dingle ratio $\tau_{tr}/\tau_q$ being considerably larger than 1 in our InSb QW samples. Correlation effects in the doping layer \cite{Efros1990} are neglected.

If scattering by charged background impurities was the dominant scattering mechanism, the dominant scattering angle $\varphi$ should, by nature, be large since the disorder potential is of short range. Again, the Dingle ratio however suggests that the dominant scattering stems from a small angle mechanism. Furthermore, with the second doping layer a considerable amount of electrons are additionally added to the QW if compared to sample C. These may neutralize charged background impurities in the well and let the remaining surplus of charged carriers contribute fully to the electron transport, hence increasing the mobility of the sample considerably. It should not be disregarded, that the lower doping layer is capable of shielding the QW from charged background impurities in the lower barrier and the bulk.

From the calculations we find, that interface roughness scattering has a minor impact on our high mobility sample E, even though the QW stands at a relatively narrow 23 nm. This comes as a surprise, as the QW width $w_0$ enters with the sixth power into the random potential [see Eq. (\ref{U_int})], which greatly increases the sensitivity to the interface roughness disorder potential as the QW gets narrower \cite{Orr2008}.

Alloy disorder is known to be prominent in quantum wells of ternary compounds \cite{Walukiewicz1984,Ogale1984,Gardner2013} where thereof induced strain additionally introduces disorder \cite{Lyo1992} or in narrow well structures with high alloy concentration in the barrier \cite{Ogale1984}. From an engineering point of view, the main factors able to influence the alloy disorder random potential are the strength parameter $V_{\textrm{AD}}$ and the AD form factor $F_{\textrm{AD}}$. With the choice of the barrier material $V_{\textrm{AD}}$ is given intrinsically. In comparison to a SSD structure, implementing a DSD structure allows to reduce the probability of the charge carrier wave function residing in the barrier material. This will reduce $F_{\textrm{AD}}$ and therefore increase the alloy scattering time $\tau_{alloy}$. To fathom the difference, we have performed $8\times8$ k$\cdot$p simulations for samples C (SSD) and E (DSD). For the lowest energy levels, the part of the total squared wave function residing in the barrier is 4.79\% (sample C) and 2.36\% (sample E), which nicely shows the effect of centering on the probability distribution of the wave function. Due to the small Al percentage in the barriers and the large Dingle ratio, we evaluate the effect of alloy scattering being of minor importance. In fact, our calculations confirm that the influence of alloy scattering is negligible in our samples. 

At room temperature (RT), the sample shows a charge carrier density of $1.04\times10^{12}$ cm$^{-2}$ with a very high mobility of 58\,000 cm$^2$/Vs, which is very close to the achievable bulk value of roughly 77\,000 cm$^2$/Vs and in range of the highest values reported \cite{Gilbertson2009}. Fig. \ref{mobplot} indicates that the RT mobility of sample E is limited by the optical phonon branch. Since in the transport model described by Davies the effective mass and the QW width are the only parameters defining $\tau_{op}$ which are sensitive to the design of the heterostructure, we assume that for a well width of 23 nm the maximum possible RT mobility is nearly reached.

%******************************************** Conclusions **********************************************%

\section{Conclusions}

In this work, we investigated the surface morphology and electron transport of a range of single- and double-side $\delta$-doped InSb quantum well heterostructures with In$_{1-x}$Al$_{x}$Sb barriers ($x=0.10$) grown on GaAs and GaSb substrates by molecular beam epitaxy, while adapting either AlSb or GaSb metamorphic buffer systems. A notable reduction of threading dislocations and hillocks down to very low values of $6.2\times10^7$ cm$^{-2}$ and $1.3\times10^7$ cm$^{-2}$, respectively, can be achieved when implementing GaSb instead of commonly used AlSb buffers on GaAs substrates. As a consequence, we observe smoother surfaces and higher electron mobilities.
Optimizing our heterostructures by additionally integrating interlayers with a higher Al content of 30\% into the InAlSb buffer, the mobility of all our samples has substantially increased. 
%This can be accredited to the high reactivity of Al adatoms, that are capable of attracting background impurities and - with our optimized buffer system - are incorporated far away from the active region.

Using a GaSb substrate smooths the sample landscape even further to a low rms roughness of 2.844 nm on $5\times5$ $\mu$m$^2$ windows, which is accompanied by a high mobility of 241\,000 cm$^2$/Vs with an electron density of $4.02\times10^{11}$ cm$^{-2}$ measured at 1.3 K for an optimized single-side $\delta$-doped sample. The fact that our samples on GaSb substrates and those grown on GaAs substrates with GaSb buffers show similar threading dislocation and hillock densities suggests, that the structural integrity and quality of our InSb quantum wells is solely determined by the growth conditions at the transition from GaSb to InAlSb and during the second intermediate buffer.

\begin{figure*}
	\centering
	\includegraphics[width=0.9\linewidth]{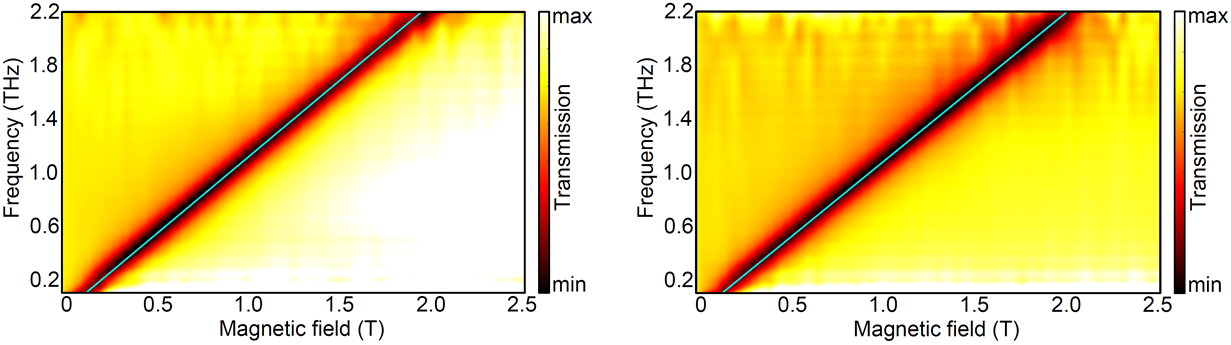}
	\caption{Transmission spectra as a function of the magnetic field showing the cyclotron resonance frequency of samples E (left) and F (right). The straight line through the resonance minimum valley represents the fit to the data for the effective mass calculation.}
	\label{cycloEF}
\end{figure*}

The highest-quality samples were achieved with double-side $\delta$-doped InSb quantum wells grown on GaAs substrates with GaSb metamorphic buffers, showing quantized Hall plateaus at zero longitudinal resistance and oscillations up to filling factors of $\nu=58$. Spin splitting at a high filling factor of $\nu=25$ is indicative of the electrons in the InSb quantum well exhibiting strong spin-orbit coupling. By applying a simple approximation, we derive a very large effective $g$-factor of $\lvert g^* \rvert =38.4$. With an electron mobility of 349\,000 cm$^2$/Vs at a density of $4.90\times10^{11}$ cm$^{-2}$ measured at 250 mK, our InSb quantum wells belong to the best quality samples reported.

Transport lifetimes indicate the dominant scattering mechanism at cryogenic temperatures to predominantly originate from ionized impurities remote to the well. This is supported by a large Dingle ratio, which for our samples is as high as $\tau_{tr}/\tau_q=37.9$. For increasing temperatures, the large-range Coulomb scattering potentials get suppressed by the influence of optical phonon scattering, limiting the mobility at room temperature to 58\,000 cm$^2$/Vs at an electron density of $1.04\times10^{12}$ cm$^{-2}$.

In addition, we show results of an inverted SSD sample with a quantum well 50 nm beneath the surface. Despite the close proximity to the surface, the sample shows a mobility of 107\,000 cm$^2$/Vs corresponding to a large electron mean free path of almost 1 $\mu$m. The sample shows a rms roughness of 0.679 nm on this length scale. Together with the very large $g^*$-factor, these results make InSb quantum wells increasingly competitive for nanoscale fabrication of Majorana based devices.

%******************************************** Acknowledgments **********************************************%

\begin{acknowledgments}
	The authors acknowledge the financial support by the Swiss National Science Foundation (SNF) through the NCCR 'QSIT-Quantum Science and Technology' (National Center of Competence in Research).
\end{acknowledgments}

%************************************************ Appendix ************************************************%
\appendix*
\section{Effective electron mass determination}

Determining the effective $g^*$-factor, the quantum scattering lifetime $\tau_q$ and several scattering rates in the transport lifetime model requires the knowledge of the effective electron mass $m^*$ of the charge carriers in the QW. Cyclotron resonance measurements in a THz time-domain spectroscopy setup allow for a precise determination of $m^*$ by
\begin{equation}
\omega = \frac{eB}{m^*}\,.
\end{equation}
Figure \ref{cycloEF} shows the cyclotron resonance data with a linear fit of samples E and F, respectively, for a magnetic field range of 0 to 2.5 T at a sample temperature of $T=3$ K. The data fitting results in effective electron masses of $m^*=0.0248 m_e\,\pm\, 0.13\%$ (sample E) and $m^*=0.0254m_e\,\pm\, 0.14\%$ (sample F). These values are in good agreement with the effective electron masses retrieved from simple $8\times8$ k$\cdot$p simulations.

\bibliography{ReferencesSummary}

\end{document}